\documentclass[article,prb,twocolumn,showpacs,superscriptaddress,nobibnotes]{revtex4-1}

\usepackage{graphicx}
\usepackage{amssymb}
\usepackage{bm}

\usepackage{amsmath}
\usepackage[dvipsnames]{xcolor}

\usepackage[colorlinks=true,
  		linkcolor=blue,
	        citecolor=blue,
	        urlcolor=blue]{hyperref}

\def\vecsign{\mathchar"017E}
\def\dvecsign{\smash{\stackon[-1.95pt]{\vecsign}{\rotatebox{180}{$\vecsign$}}}}
\def\dvec#1{\def\useanchorwidth{T}\stackon[-4.2pt]{#1}{\,\dvecsign}}
\usepackage{stackengine}
\stackMath

\newcommand{\marvel}{National Centre for Computational Design and Discovery of Novel Materials~(MARVEL), \'Ecole Polytechnique F\'ed\'erale de Lausanne, CH-1015 Lausanne, Switzerland}
\newcommand{\dqmp}{Department of Quantum Matter Physics, University of Geneva, 24 Quai Ernest Ansermet, CH-1211 Geneva, Switzerland}
\newcommand{\gap}{Group of Applied Physics, University of Geneva, 24 Quai Ernest Ansermet, CH-1211 Geneva, Switzerland}	
\newcommand{\nims}{National Institute for Materials Science, 1-1 Namiki, Tsukuba, 305-0044, Japan}

\newcommand\3{$_3$}
\newcommand{\cm}{cm\ensuremath{^{-1}}}

\newcommand{\beq}{\begin{equation}\begin{aligned}}
\newcommand{\eeq}{\end{aligned}\end{equation}}

\newcommand{\mum}{$\mu$m}

\begin{document}
\title{Low-temperature monoclinic layer stacking  in atomically thin CrI\3 crystals}
\date{\today}
\author{Nicolas Ubrig}
\email{nicolas.ubrig@unige.ch}
\affiliation{\dqmp}
\affiliation{\gap}
\author{Zhe Wang}
\affiliation{\dqmp}
\affiliation{\gap}
\author{J\'er\'emie Teyssier}
\affiliation{\dqmp}
\affiliation{\gap}
\author{Takashi Taniguchi}
\affiliation{\nims}
\author{Kenji Watanabe}
\affiliation{\nims}
\author{Enrico Giannini}
\affiliation{\dqmp}
\author{Alberto F. Morpurgo}
\affiliation{\dqmp}
\affiliation{\gap}
\author{Marco Gibertini}
\email{marco.gibertini@unige.ch}
\affiliation{\dqmp}
\affiliation{\marvel}

\begin{abstract}
Chromium triiodide, CrI\3, is emerging as a promising magnetic two-dimensional semiconductor where spins are ferromagnetically aligned within a single layer. Potential applications in spintronics arise from an antiferromagnetic ordering between adjacent layers that gives rise to spin filtering and a large magnetoresistance in tunnelling devices. This key feature appears only in thin multilayers and it is not inherited from bulk crystals, where instead neighbouring layers  share the same ferromagnetic spin orientation. This discrepancy between bulk and thin samples is unexpected, as magnetic ordering between layers arises from exchange interactions that are local in nature and should not depend strongly on thickness.  
Here we solve this controversy and show through polarization resolved Raman spectroscopy that thin multilayers do not undergo a structural phase transition typical of bulk crystals.  As a consequence, a different stacking pattern is present in thin and bulk samples at the temperatures at which magnetism sets in and, according to previous first-principles simulations, this results in a different interlayer magnetic ordering. Our experimental findings provide evidence for the strong interplay between stacking order and magnetism in CrI\3, opening interesting perspectives to design the magnetic state of van der Waals multilayers.
\end{abstract}

\maketitle

The discovery of magnetic order\cite{Huang2017,Gong2017} has disclosed novel opportunities in the field of two dimensional (2D) van der Waals crystals and heterostructures\cite{Burch_review_2018,Gong_review_2019,Gibertini_review_2019}. In many cases\cite{Gong2017,Deng2018,Fei2018}, the magnetic configuration of thin layers is the same as in the 3D parent compounds, although possibly with a reduced critical temperature owing to the larger sensitivity of 2D magnets to thermal fluctuations. This is expected as the intra- and inter-layer exchange interactions that determine the ground-state magnetic configuration are typically local and do not change significantly when thinning down the material.  

A surprising exception is represented by chromium triiodide, CrI\3, a van der Waals material that in its bulk form shows ferromagnetic (FM) ordering both within and between layers below a critical temperature $T_{\rm c} \simeq 61$~K\cite{McGuire2015,Dillon1965}. Recently, a multitude of experiments, ranging from magneto-optical Kerr effect measurements\cite{Huang2017,Song2018} to tunnelling magnetotransport\cite{Song2018,Klein2018,Wang2018,Kim2018} and scanning magnetometry\cite{Thiel2019}, have shown unarguably that instead thin samples up to at least $\sim 10$ layers display an antiferromagnetic (AFM) interlayer exchange coupling between the ferromagnetic layers. The AFM ordering can be manipulated through external electric fields\cite{Jiang2018field} or doping\cite{Jiang2018doping} and it is responsible for a spin-filtering effect on electrons tunnelling through CrI\3 barriers, giving rise to a record-high magnetoresistance\cite{Klein2018,Song2018,Wang2018,Kim2018} with potential application in spin transistors\cite{Jiang2019,Song2019}.

\begin{figure}[t]
	\centering
	\includegraphics[width=.5\textwidth]{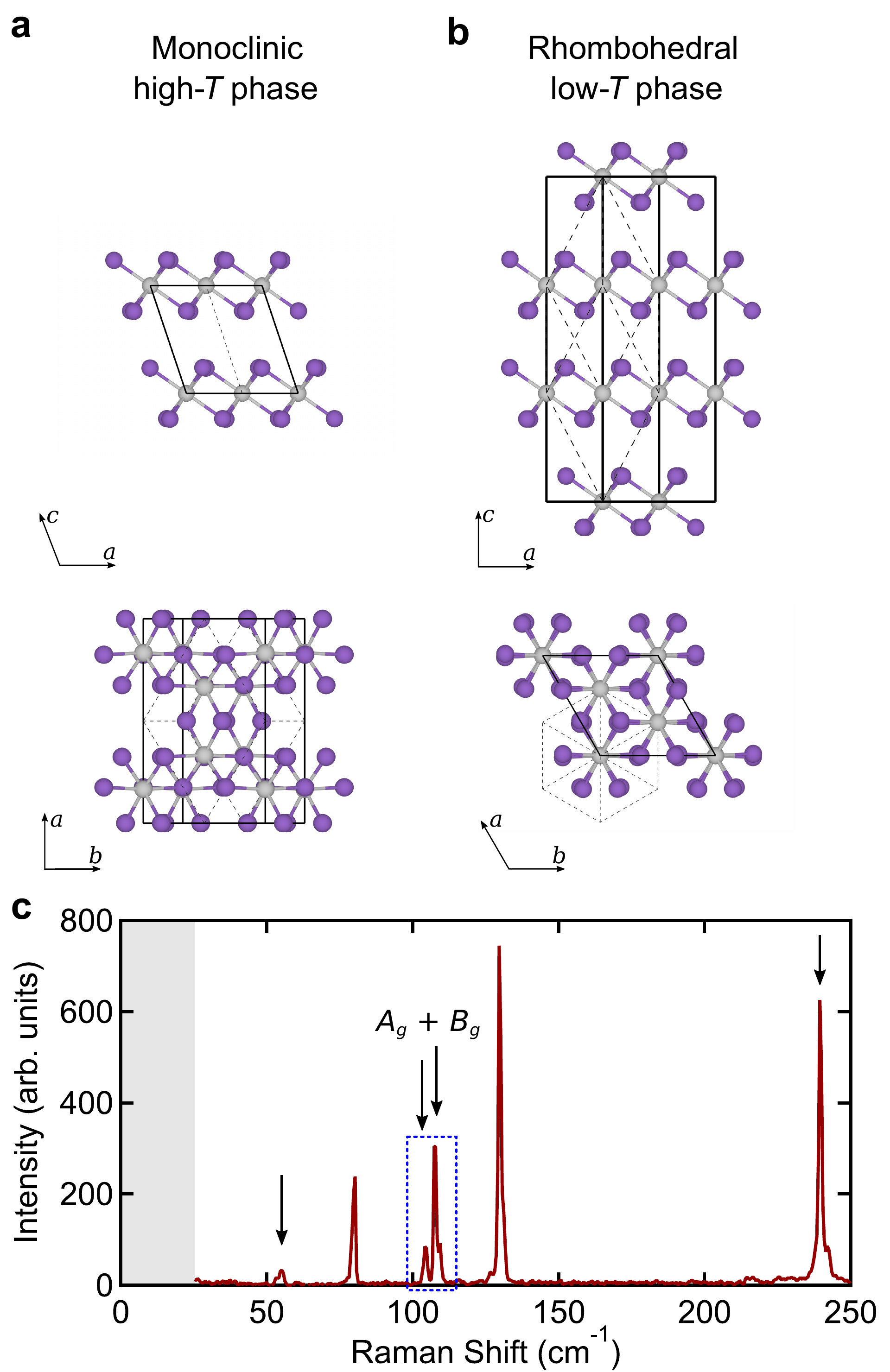}
	\caption{Lateral and top views of the crystal structure of CrI\3 in the monoclinic (high temperature) phase (\textbf{\sffamily a}) and in the rhombohedral (low temperature) phase (\textbf{\sffamily b}). Both the primitive (thin dashed line) and the conventional (thick solid line) unit cells are reported. \textbf{\sffamily c}. Typical Raman spectrum of bulk CrI\3 at room temperature without resolving the polarization. The black arrows indicate the pairs of $A_g$ and $B_g$ vibrational modes sensitive  to the structural phase transition. The blue dashed rectangle highlights the Raman active modes on which we focus our attention in the following.}
	\label{fig:01}
\end{figure}

In an attempt to clarify this unexpected change in magnetic ordering from bulk to few layers, most theoretical investigations have focused on the presence of a  structural phase transition in bulk CrI\3 at about 200-220~K \cite{McGuire2015,Djurdji2018}. Across this transition, the structure evolves from a high-temperature monoclinic phase (Fig.~\ref{fig:01}a, space group $C2/m$)  to a rhombohedral structure (Fig.~\ref{fig:01}b, space group $R\overline{3}$) at low temperature, with the main distinction  between the two phases being a different stacking order of the layers. First-principles simulations in Ref.~\onlinecite{Wang2018}, corroborated by additional theoretical investigations\cite{Sivadas2018,Soriano2018,Jang2019,Jiang2019theory,Lei2019}, have shown that the interlayer exchange coupling is FM in the rhombohedral phase (in agreement with experimental observations for bulk CrI\3), while it is AFM in the monoclinic structure. The strong interplay between stacking order and magnetic configuration suggests a possible scenario to solve the conundrum: if thin samples exfoliated at room temperature from bulk monoclinic crystals are not able to undergo a structural phase transition, they remain in the metastable monoclinic phase and are thus expected to display AFM ordering at low temperature.  

This picture is in agreement with recent measurements on few-layer CrI\3 where either an accidental puncture\cite{Thiel2019} or an external pressure\cite{Song2019pressure,Li2019} provided the energy to undergo a structural transformation with a corresponding transition to FM ordering. Additional validations supporting the connection between crystal structure and magnetism have been achieved in a related material, CrBr\3, by observing different magnetic ordering associated with novel stacking patterns (not corresponding to the bulk phases) in bilayers grown by molecular beam epitaxy\cite{Chen2019}. The ultimate confirmation of the proposed scenario requires a technique sensitive to the stacking order of few layer structures in order to verify the absence of structural phase transition in few layers. In this regard, second-harmonic generation is an effect which is sensitive to the different crystal symmetry in the two phases and that has been recently adopted\cite{Sun2019} to show that bilayer CrI\3 remains monoclinic down to very low temperature. Alternatively, another practical approach that has been successfully employed\cite{Klein2019} in a similar material, CrCl\3, relies on polarization resolved Raman spectroscopy. 

In this work, we show the absence of structural phase transitions in thin CrI\3 through polarization resolved Raman spectroscopy. Based on general symmetry arguments, we develop a strategy to distinguish the monoclinic and rhombohedral phases by looking at the angular dependence of the Raman response to linearly polarized light. We validate this approach for bulk crystals by evidencing the existence of a monoclinic phase at high temperature and a rhombohedral one at low temperature, in agreement with the general understanding. Raman measurements on encapsulated CrI\3 multilayers on the contrary show that thin crystals remain in the monoclinic phase even when the sample is cooled down to base temperature. This has crucial implications on the magnetic ground state of atomically thin samples, which is very sensitive to the stacking order of the layers, and finally explains the controversial AFM ordering observed in experiments.

\section*{Methods}

CrI\3 crystals are grown by the chemical vapor transport method and, owing to the enormous sensitivity of this material to atmosphere, stored in a nitrogen-gas-filled glove box with sub-ppm concentration of O$_2$ and H$_2$O. The investigated bulk crystals are freshly cleaved, mounted on a He-flow cryostat (cryovac KONTI cryostat) in the glove box, and sealed in the vacuum chamber with optical access before being transferred to the optical setup. The nm-thick multilayers of CrI\3 are obtained by mechanical exfoliation with scotch tape. The flakes are then picked up with standard dry transfer techniques and fully encapsulated in 10-30 nm thick exfoliated hBN. The samples are removed from the glovebox and placed into the cryostat for optical investigations.

All Raman spectroscopy measurements in this work are performed using a Horiba scientific (LabRAM HR Evolution) confocal microscope in backscattering geometry. The nominal laser power before the microscope objective and the window of the cryostat is 60 $\mu$W and the excitation wavelength 532 nm. After laser excitation the dispersed light is sent to a Czerni-Turner spectrometer equipped with a 1800 groves/mm grating, which resolves the optical spectra with a precision of 0.3 \cm{}. The light is detected with the help of a N$_2$-cooled CCD-array. The incident linear polarization of the laser is varied using a $\lambda$/2-plate while the analyzer, placed on the detecting light path, is kept fix. 

First-principles simulations have been performed within density functional theory using the Quantum ESPRESSO suite of codes\cite{Giannozzi2009,Giannozzi2017}. In order to treat magnetism and van der Waals interactions on an equal footing we have adopted the spin-polarized extension\cite{Thonhauser2015} of the van der Waals density functional (vdw-DF) method\cite{Dion2004,Lee2010}. The unit cell is kept fixed to the experimentally reported structure\cite{McGuire2015} for both the rhombohedral and monoclinic phase, while atomic positions have been relaxed so that any component of the force on any atom does not exceed $2.5\times10^{-3}$ eV/\AA. Phonon frequencies at vanishing wave vector have been then computed by finite differences using the phonopy software\cite{Togo2015}. From the computed phonon displacement patterns, the Raman tensors have been calculated within the Placzec approximation as derivatives of the electronic contribution to the dielectric tensor with respect to the phonon amplitude, again using finite differences. In all calculations, we adopt pseudopotentials from the Standard Solid State Pseudopotential Library (SSSP)\cite{Prandini2018}, with a cutoff of 60~Ry for wavefunctions and 480~Ry for the charge density. The Brillouin zone corresponding to the primitive unit cell is sampled using a regular Monkhorst-Pack grid centered at $\Gamma$ with $8\times8\times8$ or $6\times6\times6$ k-points in either the monoclinic or rhombohedral phase.

\section*{Results and discussion}

Fig.~\ref{fig:01}c shows a typical Raman spectrum of bulk CrI\3 at room temperature, in agreement with previous literature\cite{Djurdji2018}. The visible modes  belong to either one of two possible irreducible representations, $A_g$ or $B_g$, of the $2/m$  (or $C_{2h}$) point group corresponding to the high-temperature phase. When the structure undergoes a transition to the rhombohedral phase, some pairs of $A_g$ and $B_g$ modes (highlighted in Fig.~\ref{fig:01}) become degenerate and transform according to the two-dimensional $E_g$ irreducible representation of the low-temperature $\bar3$ (or $C_{3i}$) point group. The presence of degenerate or split modes thus represents a potential signature to distinguish between the two structural phases and to track the phase transition. Still, the frequency separation between the split modes is typically very small\cite{Djurdji2018} (few \cm) and thus expected to get harder to be visible in unpolarized Raman spectra when the thickness of the sample is narrowed down and the signal gets weaker. 

Such difficulty can be overcome in polarization resolved Raman spectroscopy by exploiting the different Raman response of $A_g$, $B_g$, and $E_g$ modes to polarized light. Indeed, as we shall see, the dependence on the polarization angle cancels out when the contribution from the two degenerate modes forming a  $E_g$ peak is summed over in the rhombohedral phase, resulting in a constant spectrum insensitive to the polarization configuration. On the contrary, in the monoclinic structure the $A_g$ and $B_g$ modes that result from the splitting of the degenerate $E_g$ peak have opposite response, giving rise to two peaks with an intensity that oscillates out of phase with the polarization angle, so that the close $A_g$ and $B_g$ peaks are much easier to resolve\cite{Djurdji2018,Klein2019}.

To provide a rigorous foundation of this procedure, we first recall that in non-resonant Stokes conditions, the Raman spectrum can be expressed as
\begin{align}\label{eq:intensity}
I(\omega) &= I_0\sum_{\nu} \frac{\omega_{I}\omega_{S}^{3}}{\omega_\nu} \mid {\bm\epsilon}_{S} \cdot \dvec{\bm R}^{\nu} \cdot {\bm\epsilon}_{I} \mid^{2} [n_{\nu} +1] \delta(\omega-\omega_{\nu})~,
\end{align}
where the line-shape has been simplified to a $\delta$-function, $\omega$ is the Raman shift, $\omega_{\nu}$ is the frequency of the  $\nu$-th long-wavelength phonon mode, ${\bm\epsilon}_{I}$ and ${\bm\epsilon}_{S}$ the polarization vectors of the incident and scattered light with frequency $\omega_{I}$ and $\omega_{S} = \omega_{I}-\omega$, and $n_\nu = [\exp(\hbar\omega_\nu/(k_B T))-1]^{-1}$ is the Bose-Einstein occupation of the $\nu$-th mode. Based on symmetry arguments, the Raman tensors $\dvec{\bm R}^{\nu}$ entering Eq.~\eqref{eq:intensity} have the following general expressions for modes belonging to the $A_g$ or $B_g$ representations of the high-temperature point group 
\begin{equation}
\dvec{\bm R}^{A_g} = 
\begin{pmatrix}
a & 0 & d \\ 
0 & c & 0 \\ 
d & 0 & b \\ 
\end{pmatrix}
\quad
\dvec{\bm R}^{B_g} = 
\begin{pmatrix}
0 & e & f \\ 
e & 0 & 0 \\ 
f & 0 & 0 \\ 
\end{pmatrix}
\end{equation}
and 
\begin{equation}
\dvec{\bm R}^{\phantom{.}^1E_g} = 
\begin{pmatrix}
m & n & p \\ 
n & -m & q \\ 
p & q & 0 
\end{pmatrix}
\quad
\dvec{\bm R}^{\phantom{.}^2E_g} = 
\begin{pmatrix}
n & -m & -q \\ 
-m & -n & p \\ 
-q & p & 0 
\end{pmatrix}
\end{equation}
for pairs of degenerate modes belonging to the $E_g$ representation of the low-temperature point group.

\begin{figure}
\centering
\includegraphics[width=0.95\linewidth]{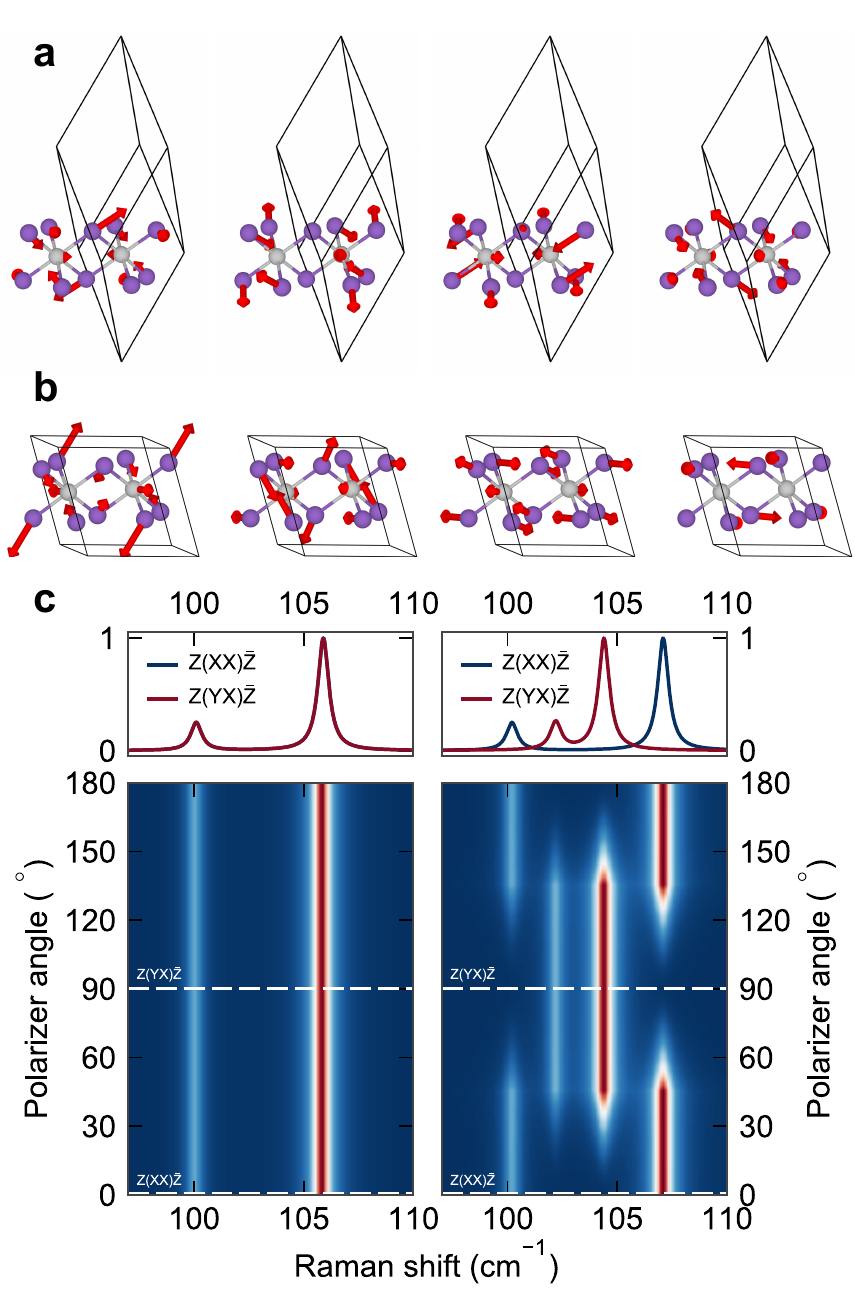}
\caption{Phonon displacement pattern according to first-principles simulations of the modes visible in the Raman spectrum close to 100~\cm{} either in the rhombohedral (\textbf{\sffamily a}) or monoclinic  (\textbf{\sffamily b}) phase. The modes appear in order of increasing frequency. \textbf{\sffamily c}: Colour plot of the normalized Raman spectrum as a function of the polarization angle and Raman shift. Results for both the rhombohedral (left) and monoclinic (right) phase of bulk CrI\3  are reported. Here we assume that the polarization angle of the incident light $\theta_I$ is varied, while keeping the detector for the scattered light fixed at $\theta_S=0$. Intensities and vibrational frequencies are computed from first principles as detailed in the Methods, although the position of the brightest $A_g$ and $B_g$ modes of the monoclinic structure (as well as the corresponding $E_g$ mode in the rhombohedral phase) have been displaced by 4~\cm{} to obtain a better qualitative agreement with experiments (see below). The upper insets show the Raman spectra in parallel (blue) or cross (red)  polarization, corresponding to the horizontal dashed lines in the colour plots.}
\label{fig:02}
\end{figure}

To identify a strategy to distinguish the two phases, we focus on the most common back-scattering geometry with linearly polarized light, whose polarization vectors can thus be written as ${\bm\epsilon}_{I/S}  = (\cos\theta_{I/S},\sin\theta_{I/S},0)$. To simplify the derivation we observe that, as the degenerate $E_g$ modes split into a $A_g$ and a $B_g$ mode, we expect $a\approx -c$ (and $|a| \approx |e|$), so that $R_{yy}^\nu\approx -R_{xx}^\nu$ for all the modes considered here. As a consequence, we find
\begin{align}
{\bm\epsilon}_{S} \cdot \dvec{\bm R}^{\nu} \cdot {\bm\epsilon}_{I}  =& \cos\theta_S\cos\theta_I R_{xx}^\nu + \sin\theta_S\sin\theta_I R_{yy}^\nu \notag\\ 
&+ \sin(\theta_S+\theta_I) R_{xy}^\nu
\\ =& \cos\theta R_{xx}^\nu + \sin\theta R_{xy}^\nu~, \notag
\end{align}
yielding an intensity in Eq.~\eqref{eq:intensity} that can be written in terms of the cumulative angle $\theta=\theta_S+\theta_I$. From the general expressions for the Raman tensors, we find that the Raman spectrum is completely independent of the polarization angles close to a degenerate $E_g$ peak in the low-temperature phase,
\begin{equation}
I_{E_g} (\omega) = I_0 \frac{\omega_{I}(\omega_{I}-\omega)^{3}}{\omega_{E_g}} [n_{E_g} +1] (m^2 + n^2)  \delta(\omega-\omega_{E_g}) 
\end{equation}
while the intensity of the split $A_g$  and $B_g$ modes of the monoclinic structure oscillates out of phase as a function of the angle $\theta$:
\begin{align}
&I_{A_g+B_g} (\omega) = I_0 \omega_{I}(\omega_{I}-\omega)^{3} a^2 \times  \\
&\bigg[ \frac{n_{A_g} +1}{\omega_{A_g}}  \cos^2\theta  \delta(\omega-\omega_{A_g})  +  \frac{n_{B_g} +1}{\omega_{B_g}}  \sin^2\theta   \delta(\omega-\omega_{B_g}) \bigg]~.\notag
\end{align}

\begin{figure*}
\centering
\includegraphics[width=\textwidth]{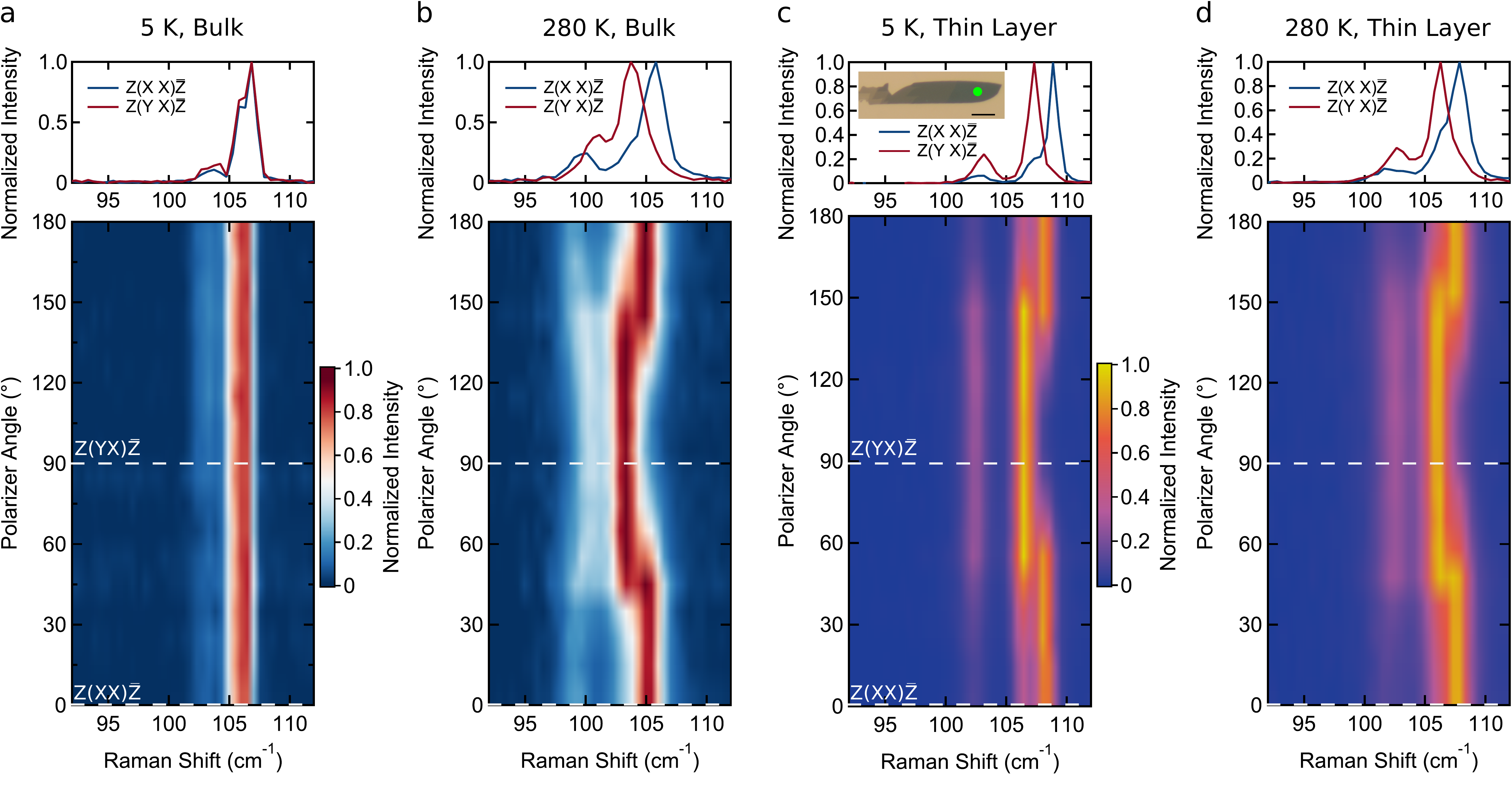}
\caption{Color plots of the normalized intensity as a function of the Raman shift (in \cm{}) in the investigated frequency region and of the angle of the incoming linearly polarized light. The white dashed lines indicate co- and cross-polarized configurations ($\mathsf{Z(XX)\Bar{Z}}$ and $\mathsf{Z(YX)\Bar{Z}}$, respectively). The top panels show individual line cuts for $\mathsf{Z(XX)\Bar{Z}}$ and $\mathsf{Z(YX)\Bar{Z}}$. Bulk CrI\3 at $T = 5$~K (\textbf{\sffamily a}) and $T = 280$~K (\textbf{\sffamily b}) show excellent agreement with the theoretical predictions showing that the structural phase transition from monoclinic, at high temperatures, to rhombohedral, at low temperatures, takes place. The experimental data for a 4~nm thick crystal of CrI\3 is shown in \textbf{\sffamily c} and \textbf{\sffamily d} for $T = 5$~K and $T =280$~K, respectively. The area probed here and an optical micrograph of the flake is depicted in the inset of panel \textbf{\sffamily c}. The scale bar corresponds to 10 \mum{}. Except for a slight stiffening of the modes due to the decreased temperature we do not observe significant differences between the experimental data at low and high temperatures. In particular, the polarization pattern is virtually identical, indicating that no structural transition occurs so that the system remains in the monoclinic phase down to low temperature.}
\label{fig:03}
\end{figure*}

This provides us with a strategy to distinguish the two structural phases by looking at the evolution of the Raman spectrum as the cumulative polarization angle $\theta$ is varied, e.g. by keeping fixed $\theta_{S}$ while sweeping $\theta_I$. In the rhombohedral phase we expect degenerate $E_g$ peaks whose intensity and frequency position do not change with the polarization angle, while the monoclinic structure is characterized by pairs of close peaks associated with $A_g$ and $B_g$ modes whose intensity alternates out of phase as a function of $\theta$, enhancing the visibility of the split modes even when the frequency separation is very small. In the following we focus on the   spectral range around 100~\cm, where this strategy is particularly suitable owing to the presence\cite{Djurdji2018} of two nearby $E_g$ modes in the rhombohedral phase split into two pairs of $A_g+B_g$ modes in the monoclinic phase. 

The procedure is exemplified in Fig.~\ref{fig:02} in a model calculation, where the Raman intensities and vibrational frequencies have been computed for the rhombohedral and monoclinic structures from first principles (see Methods), although the position of the strongest $A_g+B_g$ peaks of the monoclinic crystal (and for consistency also the corresponding $E_g$ peak in the rhombohedral phase) have been shifted by 4~\cm{} to result in a better qualitative agreement with experiments. The angular dependence of the two spectra are clearly different, supporting the effectiveness of our strategy to distinguish the two structural phases. Indeed, as expected from the above discussion, for the rhombohedral structure, the Raman spectrum is insensitive to the polarization angle $\theta_I$ (here we assume $\theta_S=0$), while in the monoclinic phase we have a transfer of spectral intensity as a function of $\theta_I$ between two peaks of a split $A_g+B_g$ pair. In particular, this results into very different spectra in parallel configuration ($\theta_I=\theta_S=0$, or $\mathsf{Z(XX)\bar{Z}}$ in Porto notation), where only $A_g$ modes are visible, or cross configuration ($\theta_I=\pi/2$, or $\mathsf{Z(YX)\bar{Z}}$), where only $B_g$ modes are present, allowing to clearly resolve their frequency separation. 

We first validate our approach by considering bulk samples, for which a structural phase transition is expected to occur at 200-220~K and should manifest itself in a change of the angular dependence of the Raman spectrum. Fig.~\ref{fig:03} shows the Raman response measured at two different temperatures, above and below the structural phase transition, as a function of the polarization angle $\theta_I$ of the incident light, while the detecting polarizer is kept fixed (see Methods).  At 5~K a weak and a strong peak are present at 103 and 107~\cm{} respectively, and their intensity does not evolve with $\theta_I$, clearly showing that these are $E_g$ modes of the rhombohedral structure. 

In particular, the relative intensity of the two modes is in very good agreement with the first-principles results in Fig.~\ref{fig:02} and allows us to make a more definite assignment of the corresponding phonon patterns. At high temperature (280~K), each peak splits into two with intensities that oscillate out of phase as a function of $\theta_I$, so that for parallel and cross polarization (upper panel) only one of the two split modes is visible. This is exactly what is predicted for a monoclinic structure in Fig.~\ref{fig:02} and we can thus unambiguously identify the peaks as pairs of $A_g+B_g$ modes, indicating that the  structure is monoclinic at high temperature.  Our approach thus confirms that bulk crystals undergo a structural transition from a monoclinic phase at high temperature to a rhombohedral phase at low temperature. 

We are now in a position to consider thin samples obtained by mechanical exfoliation (see Methods). Fig.~\ref{fig:03} shows the Raman spectra obtained at 5 and 280~K for a 4~nm thick CrI\3 crystal (see inset, approximately 6 layers) as a function of the polarization angle. For definiteness, we focus on the same spectral range considered for bulk samples. In this case, apart from a clear reduction of the peak width upon decreasing temperature, the two spectra are virtually identical. In particular, both at high and low temperature we find two pairs of close by peaks whose intensity varies with the polarization angle in phase opposition. Such transfer of spectral intensity as a function of $\theta_I$ between nearby peaks is the clear signature of the monoclinic phase introduced before, ruling out the emergence of the rhombohedral phase.

As temperature plays a crucial role in rhombohedral-monoclinic transition, sound conclusions on the presence or absence of structural changes require an independent cross check of the effective temperature of the sample for every experiment. Indeed, in a measurement of the Raman spectrum of insulators --which usually have low thermal conductivity-- the laser can heat up the illuminated sample area. For instance the temperature can be lifted locally above a phase transition temperature, potentially leading to spectra that artificially look similar even at very different nominal temperatures\cite{Klein2019}. Through the analysis of the intensity ratio between the Stokes and the anti-Stokes peaks in the entire spectral range\cite{mcgrane_quantitative_2014} we ensure that the sample area which is probed remains below the temperatures of the phase transitions.

We can thus safely state that thin samples remain in the monoclinic structure down to very low temperature, even below the critical temperature $T_c$ at which magnetism sets in. In this respect, the persistence of the monoclinic phase explains the observation\cite{Song2018,Klein2018,Wang2018,Kim2018} of layer antiferromagnetism in thin crystals, as opposed to the bulk FM order. Indeed, the monoclinic stacking order has been predicted to favour an AFM interlayer exchange coupling according to density-functional-theory simulations\cite{Wang2018,Sivadas2018,Soriano2018,Jang2019,Jiang2019theory,Lei2019}. The different magnetic state also results in a change in critical temperature, from 61~K in the bulk to 51~K (Ref.~\onlinecite{Wang2018}) in thin crystals. 

Remarkably, this reduced $T_c$ matches exactly the temperature at which an anomaly is observed\cite{McGuire2015,Wang2018} in the magnetization curves of bulk CrI\3. This could indicate that also the outermost layers of bulk samples do not undergo a structural transition, in the same way as thin crystals. Indeed, by remaining in the monoclinic phase, such layers would display AFM order, giving rise to an anomaly at the onset of antiferromagnetism in monoclinic CrI\3 (51~K instead of 61~K), which corresponds to the temperature of the anomaly observed in experiments. 

The common behaviour of thin crystals and the outermost layers of bulk CrI\3 would then suggest the importance of free surfaces in the suppression of the structural transition. Indeed, the absence of neighbouring layers at the surface could affect both the thermodynamics and the kinetics of the phase transition, e.g. by changing the vibrational free energy or the barrier height. Such effects could extend quite deeply inside the material or for relatively large thicknesses. Although this seems promising, further studies will be needed to clarify the precise nature and the spatial extension of the surface effects on the structural transition.

\section*{Conclusion}

In conclusion, we have identified a strategy to distinguish between the two structural phases of CrI\3 through polarization resolved Raman spectroscopy. We have validated our approach in the case of bulk crystals, confirming the existence of a structural transition from a monoclinic phase at high temperature to a rhombohedral phase at low temperature. When considering thin samples, our Raman spectroscopy analysis shows that the monoclinic structure persists down to very low temperature, clearly indicating the absence of any structural change when the thickness of the material is narrowed to few atomic layers. These results provide fundamental insight to confirm a plausible scenario that explains the full set of experimental data on CrI\3, possibly including the presence of anomalies in the magnetization curves of bulk crystals. 

\noindent {\it Note}: during the preparation of this manuscript we became aware that Raman results  similar to the ones reported here have very recently appeared in Ref.~\onlinecite{Li2019}.

\section*{Acknowledgements}
We sincerely acknowledge Alexandre Ferreira for technical support. A.F.M. gratefully acknowledges financial support from the Swiss National Science Foundation (Division II) and from the EU Graphene Flagship project. M.G.\ acknowledges support from the Swiss National Science Foundation through the Ambizione program. Simulation time was provided by CSCS on Piz Daint (project IDs s825 and s917).  K.W. and T.T. acknowledge support from the Elemental Strategy Initiative conducted by the MEXT, Japan, A3 Foresight by JSPS and the CREST (JPMJCR15F3), JST.


\begin{thebibliography}{10}%
\makeatletter
\providecommand \@ifxundefined [1]{%
 \ifx #1\undefined \expandafter \@firstoftwo
 \else \expandafter \@secondoftwo
\fi
}%
\providecommand \@ifnum [1]{%
 \ifnum #1\expandafter \@firstoftwo
 \else \expandafter \@secondoftwo
\fi
}%
\providecommand \enquote [1]{``#1''}%
\providecommand \bibnamefont  [1]{#1}%
\providecommand \bibfnamefont [1]{#1}%
\providecommand \citenamefont [1]{#1}%
\providecommand\href[0]{\@sanitize\@href}%
\providecommand\@href[1]{\endgroup\@@startlink{#1}\endgroup\@@href}%
\providecommand\@@href[1]{#1\@@endlink}%
\providecommand \@sanitize [0]{\begingroup\catcode`\&12\catcode`\#12\relax}%
\@ifxundefined \pdfoutput {\@firstoftwo}{%
 \@ifnum{\z@=\pdfoutput}{\@firstoftwo}{\@secondoftwo}%
}{%
 \providecommand\@@startlink[1]{\leavevmode}%
 \providecommand\@@endlink[0]{}%
}{%
 \providecommand\@@startlink[1]{%
  \leavevmode
  \pdfstartlink
   attr{/Border[0 0 1 ]/H/I/C[0 1 1]}%
   user{/Subtype/Link/A<</Type/Action/S/URI/URI(#1)>>}%
  \relax
 }%
 \providecommand\@@endlink[0]{\pdfendlink}%
}%
\providecommand \url  [0]{\begingroup\@sanitize \@url }%
\providecommand \@url [1]{\endgroup\@href {#1}{\urlprefix}}%
\providecommand \urlprefix [0]{URL }%
\providecommand \Eprint[0]{\href }%
\@ifxundefined \urlstyle {%
  \providecommand \doi [1]{doi:\discretionary{}{}{}#1}%
}{%
  \providecommand \doi [0]{doi:\discretionary{}{}{}\begingroup
  \urlstyle{rm}\Url }%
}%
\providecommand \doibase [0]{http://dx.doi.org/}%
\providecommand \Doi[1]{\href{\doibase#1}}%
\providecommand \bibAnnote [3]{%
  \BibitemShut{#1}%
  \begin{quotation}\noindent
    \textsc{Key:}\ #2\\\textsc{Annotation:}\ #3%
  \end{quotation}%
}%
\providecommand \bibAnnoteFile [2]{%
  \IfFileExists{#2}{\bibAnnote {#1} {#2} {\input{#2}}}{}%
}%
\providecommand \typeout [0]{\immediate \write \m@ne }%
\providecommand \selectlanguage [0]{\@gobble}%
\providecommand \bibinfo [0]{\@secondoftwo}%
\providecommand \bibfield [0]{\@secondoftwo}%
\providecommand \translation [1]{[#1]}%
\providecommand \BibitemOpen[0]{}%
\providecommand \bibitemStop [0]{}%
\providecommand \bibitemNoStop [0]{.\EOS\space}%
\providecommand \EOS [0]{\spacefactor3000\relax}%
\providecommand \BibitemShut [1]{\csname bibitem#1\endcsname}%
\bibitem{Huang2017}%
  \BibitemOpen
  \bibfield{author}{%
  \bibinfo {author} {\bibfnamefont{B.}~\bibnamefont{Huang}}, \bibinfo {author}
  {\bibfnamefont{G.}~\bibnamefont{Clark}}, \bibinfo {author}
  {\bibfnamefont{E.}~\bibnamefont{Navarro-Moratalla}}, \bibinfo {author}
  {\bibfnamefont{D.~R.}\ \bibnamefont{Klein}}, \bibinfo {author}
  {\bibfnamefont{R.}~\bibnamefont{Cheng}}, \bibinfo {author}
  {\bibfnamefont{K.~L.}\ \bibnamefont{Seyler}}, \bibinfo {author}
  {\bibfnamefont{D.}~\bibnamefont{Zhong}}, \bibinfo {author}
  {\bibfnamefont{E.}~\bibnamefont{Schmidgall}}, \bibinfo {author}
  {\bibfnamefont{M.~A.}\ \bibnamefont{McGuire}}, \bibinfo {author}
  {\bibfnamefont{D.~H.}\ \bibnamefont{Cobden}}, \bibinfo {author}
  {\bibfnamefont{W.}~\bibnamefont{Yao}}, \bibinfo {author}
  {\bibfnamefont{D.}~\bibnamefont{Xiao}}, \bibinfo {author}
  {\bibfnamefont{P.}~\bibnamefont{Jarillo-Herrero}},\ and\ \bibinfo {author}
  {\bibfnamefont{X.}~\bibnamefont{Xu}},\ }%
  \Doi{10.1038/nature22391}{\emph{\bibinfo {title} {Layer-dependent
  ferromagnetism in a van der {Waals} crystal down to the monolayer limit}}},\
  \bibinfo {journal} {Nature}\ \textbf{\bibinfo {volume} {546}},\ \bibinfo
  {pages} {270} (\bibinfo {year} {2017}).~%
  \bibAnnoteFile{Stop}{Huang2017}%
\bibitem{Gong2017}%
  \BibitemOpen
  \bibfield{author}{%
  \bibinfo {author} {\bibfnamefont{C.}~\bibnamefont{Gong}}, \bibinfo {author}
  {\bibfnamefont{L.}~\bibnamefont{Li}}, \bibinfo {author}
  {\bibfnamefont{Z.}~\bibnamefont{Li}}, \bibinfo {author}
  {\bibfnamefont{H.}~\bibnamefont{Ji}}, \bibinfo {author}
  {\bibfnamefont{A.}~\bibnamefont{Stern}}, \bibinfo {author}
  {\bibfnamefont{Y.}~\bibnamefont{Xia}}, \bibinfo {author}
  {\bibfnamefont{T.}~\bibnamefont{Cao}}, \bibinfo {author}
  {\bibfnamefont{W.}~\bibnamefont{Bao}}, \bibinfo {author}
  {\bibfnamefont{C.}~\bibnamefont{Wang}}, \bibinfo {author}
  {\bibfnamefont{Y.}~\bibnamefont{Wang}}, \bibinfo {author}
  {\bibfnamefont{Z.~Q.}\ \bibnamefont{Qiu}}, \bibinfo {author}
  {\bibfnamefont{R.~J.}\ \bibnamefont{Cava}}, \bibinfo {author}
  {\bibfnamefont{S.~G.}\ \bibnamefont{Louie}}, \bibinfo {author}
  {\bibfnamefont{J.}~\bibnamefont{Xia}},\ and\ \bibinfo {author}
  {\bibfnamefont{X.}~\bibnamefont{Zhang}},\ }%
  \Doi{10.1038/nature22060}{\emph{\bibinfo {title} {Discovery of intrinsic
  ferromagnetism in two-dimensional van der {Waals} crystals}}},\ \bibinfo
  {journal} {Nature}\ \textbf{\bibinfo {volume} {546}},\ \bibinfo {pages} {265}
  (\bibinfo {year} {2017}).~%
  \bibAnnoteFile{Stop}{Gong2017}%
\bibitem{Burch_review_2018}%
  \BibitemOpen
  \bibfield{author}{%
  \bibinfo {author} {\bibfnamefont{K.~S.}\ \bibnamefont{Burch}}, \bibinfo
  {author} {\bibfnamefont{D.}~\bibnamefont{Mandrus}},\ and\ \bibinfo {author}
  {\bibfnamefont{J.-G.}\ \bibnamefont{Park}},\ }%
  \Doi{10.1038/s41586-018-0631-z}{\emph{\bibinfo {title} {Magnetism in
  two-dimensional van der {Waals} materials}}},\ \bibinfo {journal} {Nature}\
  \textbf{\bibinfo {volume} {563}},\ \bibinfo {pages} {47} (\bibinfo {year}
  {2018}).~%
  \bibAnnoteFile{Stop}{Burch_review_2018}%
\bibitem{Gong_review_2019}%
  \BibitemOpen
  \bibfield{author}{%
  \bibinfo {author} {\bibfnamefont{C.}~\bibnamefont{Gong}}\ and\ \bibinfo
  {author} {\bibfnamefont{X.}~\bibnamefont{Zhang}},\ }%
  \Doi{10.1126/science.aav4450}{\emph{\bibinfo {title} {Two-dimensional
  magnetic crystals and emergent heterostructure devices}}},\ \bibinfo
  {journal} {Science}\ \textbf{\bibinfo {volume} {363}} (\bibinfo {year}
  {2019}).~%
  \bibAnnoteFile{Stop}{Gong_review_2019}%
\bibitem{Gibertini_review_2019}%
  \BibitemOpen
  \bibfield{author}{%
  \bibinfo {author} {\bibfnamefont{M.}~\bibnamefont{Gibertini}}, \bibinfo
  {author} {\bibfnamefont{M.}~\bibnamefont{Koperski}}, \bibinfo {author}
  {\bibfnamefont{A.~F.}\ \bibnamefont{Morpurgo}},\ and\ \bibinfo {author}
  {\bibfnamefont{K.~S.}\ \bibnamefont{Novoselov}},\ }%
  \Doi{10.1038/s41565-019-0438-6}{\emph{\bibinfo {title} {Magnetic {2D}
  materials and heterostructures}}},\ \bibinfo {journal} {Nature
  Nanotechnology}\ \textbf{\bibinfo {volume} {14}},\ \bibinfo {pages} {408}
  (\bibinfo {year} {2019}).~%
  \bibAnnoteFile{Stop}{Gibertini_review_2019}%
\bibitem{Deng2018}%
  \BibitemOpen
  \bibfield{author}{%
  \bibinfo {author} {\bibfnamefont{Y.}~\bibnamefont{Deng}}, \bibinfo {author}
  {\bibfnamefont{Y.}~\bibnamefont{Yu}}, \bibinfo {author}
  {\bibfnamefont{Y.}~\bibnamefont{Song}}, \bibinfo {author}
  {\bibfnamefont{J.}~\bibnamefont{Zhang}}, \bibinfo {author}
  {\bibfnamefont{N.~Z.}\ \bibnamefont{Wang}}, \bibinfo {author}
  {\bibfnamefont{Z.}~\bibnamefont{Sun}}, \bibinfo {author}
  {\bibfnamefont{Y.}~\bibnamefont{Yi}}, \bibinfo {author}
  {\bibfnamefont{Y.~Z.}\ \bibnamefont{Wu}}, \bibinfo {author}
  {\bibfnamefont{S.}~\bibnamefont{Wu}}, \bibinfo {author}
  {\bibfnamefont{J.}~\bibnamefont{Zhu}}, \bibinfo {author}
  {\bibfnamefont{J.}~\bibnamefont{Wang}}, \bibinfo {author}
  {\bibfnamefont{X.~H.}\ \bibnamefont{Chen}},\ and\ \bibinfo {author}
  {\bibfnamefont{Y.}~\bibnamefont{Zhang}},\ }%
  \Doi{10.1038/s41586-018-0626-9}{\emph{\bibinfo {title} {Gate-tunable
  room-temperature ferromagnetism in two-dimensional {Fe$_3$GeTe$_2$}}}},\
  \bibinfo {journal} {Nature}\ \textbf{\bibinfo {volume} {563}},\ \bibinfo
  {pages} {94} (\bibinfo {year} {2018}).~%
  \bibAnnoteFile{Stop}{Deng2018}%
\bibitem{Fei2018}%
  \BibitemOpen
  \bibfield{author}{%
  \bibinfo {author} {\bibfnamefont{Z.}~\bibnamefont{Fei}}, \bibinfo {author}
  {\bibfnamefont{B.}~\bibnamefont{Huang}}, \bibinfo {author}
  {\bibfnamefont{P.}~\bibnamefont{Malinowski}}, \bibinfo {author}
  {\bibfnamefont{W.}~\bibnamefont{Wang}}, \bibinfo {author}
  {\bibfnamefont{T.}~\bibnamefont{Song}}, \bibinfo {author}
  {\bibfnamefont{J.}~\bibnamefont{Sanchez}}, \bibinfo {author}
  {\bibfnamefont{W.}~\bibnamefont{Yao}}, \bibinfo {author}
  {\bibfnamefont{D.}~\bibnamefont{Xiao}}, \bibinfo {author}
  {\bibfnamefont{X.}~\bibnamefont{Zhu}}, \bibinfo {author}
  {\bibfnamefont{A.~F.}\ \bibnamefont{May}}, \bibinfo {author}
  {\bibfnamefont{W.}~\bibnamefont{Wu}}, \bibinfo {author}
  {\bibfnamefont{D.~H.}\ \bibnamefont{Cobden}}, \bibinfo {author}
  {\bibfnamefont{J.-H.}\ \bibnamefont{Chu}},\ and\ \bibinfo {author}
  {\bibfnamefont{X.}~\bibnamefont{Xu}},\ }%
  \Doi{10.1038/s41563-018-0149-7}{\emph{\bibinfo {title} {Two-dimensional
  itinerant ferromagnetism in atomically thin {Fe$_3$GeTe$_2$}}}},\ \bibinfo
  {journal} {Nature Materials}\ \textbf{\bibinfo {volume} {17}},\ \bibinfo
  {pages} {778} (\bibinfo {year} {2018}).~%
  \bibAnnoteFile{Stop}{Fei2018}%
\bibitem{McGuire2015}%
  \BibitemOpen
  \bibfield{author}{%
  \bibinfo {author} {\bibfnamefont{M.~A.}\ \bibnamefont{McGuire}}, \bibinfo
  {author} {\bibfnamefont{H.}~\bibnamefont{Dixit}}, \bibinfo {author}
  {\bibfnamefont{V.~R.}\ \bibnamefont{Cooper}},\ and\ \bibinfo {author}
  {\bibfnamefont{B.~C.}\ \bibnamefont{Sales}},\ }%
  \Doi{10.1021/cm504242t}{\emph{\bibinfo {title} {Coupling of Crystal Structure
  and Magnetism in the Layered, Ferromagnetic Insulator CrI3}}},\ \bibinfo
  {journal} {Chemistry of Materials}\ \textbf{\bibinfo {volume} {27}},\
  \bibinfo {pages} {612} (\bibinfo {year} {2015}).~%
  \bibAnnoteFile{Stop}{McGuire2015}%
\bibitem{Dillon1965}%
  \BibitemOpen
  \bibfield{author}{%
  \bibinfo {author} {\bibfnamefont{J.~F.}\ \bibnamefont{Dillon}}\ and\ \bibinfo
  {author} {\bibfnamefont{C.~E.}\ \bibnamefont{Olson}},\ }%
  \Doi{10.1063/1.1714194}{\emph{\bibinfo {title} {Magnetization, Resonance, and
  Optical Properties of the Ferromagnet CrI$_3$}}},\ \bibinfo {journal}
  {Journal of Applied Physics}\ \textbf{\bibinfo {volume} {36}},\ \bibinfo
  {pages} {1259} (\bibinfo {year} {1965}).~%
  \bibAnnoteFile{Stop}{Dillon1965}%
\bibitem{Song2018}%
  \BibitemOpen
  \bibfield{author}{%
  \bibinfo {author} {\bibfnamefont{T.}~\bibnamefont{Song}}, \bibinfo {author}
  {\bibfnamefont{X.}~\bibnamefont{Cai}}, \bibinfo {author}
  {\bibfnamefont{M.~W.-Y.}\ \bibnamefont{Tu}}, \bibinfo {author}
  {\bibfnamefont{X.}~\bibnamefont{Zhang}}, \bibinfo {author}
  {\bibfnamefont{B.}~\bibnamefont{Huang}}, \bibinfo {author}
  {\bibfnamefont{N.~P.}\ \bibnamefont{Wilson}}, \bibinfo {author}
  {\bibfnamefont{K.~L.}\ \bibnamefont{Seyler}}, \bibinfo {author}
  {\bibfnamefont{L.}~\bibnamefont{Zhu}}, \bibinfo {author}
  {\bibfnamefont{T.}~\bibnamefont{Taniguchi}}, \bibinfo {author}
  {\bibfnamefont{K.}~\bibnamefont{Watanabe}}, \bibinfo {author}
  {\bibfnamefont{M.~A.}\ \bibnamefont{McGuire}}, \bibinfo {author}
  {\bibfnamefont{D.~H.}\ \bibnamefont{Cobden}}, \bibinfo {author}
  {\bibfnamefont{D.}~\bibnamefont{Xiao}}, \bibinfo {author}
  {\bibfnamefont{W.}~\bibnamefont{Yao}},\ and\ \bibinfo {author}
  {\bibfnamefont{X.}~\bibnamefont{Xu}},\ }%
  \Doi{10.1126/science.aar4851}{\emph{\bibinfo {title} {Giant tunneling
  magnetoresistance in spin-filter van der {Waals} heterostructures}}},\
  \bibinfo {journal} {Science}\ \textbf{\bibinfo {volume} {360}},\ \bibinfo
  {pages} {1214} (\bibinfo {year} {2018}).~%
  \bibAnnoteFile{Stop}{Song2018}%
\bibitem{Klein2018}%
  \BibitemOpen
  \bibfield{author}{%
  \bibinfo {author} {\bibfnamefont{D.~R.}\ \bibnamefont{Klein}}, \bibinfo
  {author} {\bibfnamefont{D.}~\bibnamefont{MacNeill}}, \bibinfo {author}
  {\bibfnamefont{J.~L.}\ \bibnamefont{Lado}}, \bibinfo {author}
  {\bibfnamefont{D.}~\bibnamefont{Soriano}}, \bibinfo {author}
  {\bibfnamefont{E.}~\bibnamefont{Navarro-Moratalla}}, \bibinfo {author}
  {\bibfnamefont{K.}~\bibnamefont{Watanabe}}, \bibinfo {author}
  {\bibfnamefont{T.}~\bibnamefont{Taniguchi}}, \bibinfo {author}
  {\bibfnamefont{S.}~\bibnamefont{Manni}}, \bibinfo {author}
  {\bibfnamefont{P.}~\bibnamefont{Canfield}}, \bibinfo {author}
  {\bibfnamefont{J.}~\bibnamefont{Fern{\'a}ndez-Rossier}},\ and\ \bibinfo
  {author} {\bibfnamefont{P.}~\bibnamefont{Jarillo-Herrero}},\ }%
  \Doi{10.1126/science.aar3617}{\emph{\bibinfo {title} {{Probing magnetism in
  2D van der Waals crystalline insulators via electron tunneling}}}},\ \bibinfo
  {journal} {Science}\ \textbf{\bibinfo {volume} {360}},\ \bibinfo {pages}
  {1218} (\bibinfo {year} {2018}).~%
  \bibAnnoteFile{Stop}{Klein2018}%
\bibitem{Wang2018}%
  \BibitemOpen
  \bibfield{author}{%
  \bibinfo {author} {\bibfnamefont{Z.}~\bibnamefont{Wang}}, \bibinfo {author}
  {\bibfnamefont{I.}~\bibnamefont{Guti{\'e}rrez-Lezama}}, \bibinfo {author}
  {\bibfnamefont{N.}~\bibnamefont{Ubrig}}, \bibinfo {author}
  {\bibfnamefont{M.}~\bibnamefont{Kroner}}, \bibinfo {author}
  {\bibfnamefont{M.}~\bibnamefont{Gibertini}}, \bibinfo {author}
  {\bibfnamefont{T.}~\bibnamefont{Taniguchi}}, \bibinfo {author}
  {\bibfnamefont{K.}~\bibnamefont{Watanabe}}, \bibinfo {author}
  {\bibfnamefont{A.}~\bibnamefont{Imamo{\u g}lu}}, \bibinfo {author}
  {\bibfnamefont{E.}~\bibnamefont{Giannini}},\ and\ \bibinfo {author}
  {\bibfnamefont{A.~F.}\ \bibnamefont{Morpurgo}},\ }%
  \Doi{10.1038/s41467-018-04953-8}{\emph{\bibinfo {title} {Very large tunneling
  magnetoresistance in layered magnetic semiconductor {CrI$_3$}}}},\ \bibinfo
  {journal} {Nature Communications}\ \textbf{\bibinfo {volume} {9}},\ \bibinfo
  {pages} {2516} (\bibinfo {year} {2018}).~%
  \bibAnnoteFile{Stop}{Wang2018}%
\bibitem{Kim2018}%
  \BibitemOpen
  \bibfield{author}{%
  \bibinfo {author} {\bibfnamefont{H.~H.}\ \bibnamefont{Kim}}, \bibinfo
  {author} {\bibfnamefont{B.}~\bibnamefont{Yang}}, \bibinfo {author}
  {\bibfnamefont{T.}~\bibnamefont{Patel}}, \bibinfo {author}
  {\bibfnamefont{F.}~\bibnamefont{Sfigakis}}, \bibinfo {author}
  {\bibfnamefont{C.}~\bibnamefont{Li}}, \bibinfo {author}
  {\bibfnamefont{S.}~\bibnamefont{Tian}}, \bibinfo {author}
  {\bibfnamefont{H.}~\bibnamefont{Lei}},\ and\ \bibinfo {author}
  {\bibfnamefont{A.~W.}\ \bibnamefont{Tsen}},\ }%
  \Doi{10.1021/acs.nanolett.8b01552}{\emph{\bibinfo {title} {One Million
  Percent Tunnel Magnetoresistance in a Magnetic van der {Waals}
  Heterostructure}}},\ \bibinfo {journal} {Nano Letters}\ \textbf{\bibinfo
  {volume} {18}},\ \bibinfo {pages} {4885} (\bibinfo {year} {2018}).~%
  \bibAnnoteFile{Stop}{Kim2018}%
\bibitem{Thiel2019}%
  \BibitemOpen
  \bibfield{author}{%
  \bibinfo {author} {\bibfnamefont{L.}~\bibnamefont{Thiel}}, \bibinfo {author}
  {\bibfnamefont{Z.}~\bibnamefont{Wang}}, \bibinfo {author}
  {\bibfnamefont{M.~A.}\ \bibnamefont{Tschudin}}, \bibinfo {author}
  {\bibfnamefont{D.}~\bibnamefont{Rohner}}, \bibinfo {author}
  {\bibfnamefont{I.}~\bibnamefont{Guti{\'e}rrez-Lezama}}, \bibinfo {author}
  {\bibfnamefont{N.}~\bibnamefont{Ubrig}}, \bibinfo {author}
  {\bibfnamefont{M.}~\bibnamefont{Gibertini}}, \bibinfo {author}
  {\bibfnamefont{E.}~\bibnamefont{Giannini}}, \bibinfo {author}
  {\bibfnamefont{A.~F.}\ \bibnamefont{Morpurgo}},\ and\ \bibinfo {author}
  {\bibfnamefont{P.}~\bibnamefont{Maletinsky}},\ }%
  \Doi{10.1126/science.aav6926}{\emph{\bibinfo {title} {Probing magnetism in 2D
  materials at the nanoscale with single-spin microscopy}}},\ \bibinfo
  {journal} {Science}\ \textbf{\bibinfo {volume} {364}},\ \bibinfo {pages}
  {973} (\bibinfo {year} {2019}).~%
  \bibAnnoteFile{Stop}{Thiel2019}%
\bibitem{Jiang2018field}%
  \BibitemOpen
  \bibfield{author}{%
  \bibinfo {author} {\bibfnamefont{S.}~\bibnamefont{Jiang}}, \bibinfo {author}
  {\bibfnamefont{J.}~\bibnamefont{Shan}},\ and\ \bibinfo {author}
  {\bibfnamefont{K.~F.}\ \bibnamefont{Mak}},\ }%
  \Doi{10.1038/s41563-018-0040-6}{\emph{\bibinfo {title} {Electric-field
  switching of two-dimensional van der {Waals} magnets}}},\ \bibinfo {journal}
  {Nature Materials}\ \textbf{\bibinfo {volume} {17}},\ \bibinfo {pages} {406}
  (\bibinfo {year} {2018}).~%
  \bibAnnoteFile{Stop}{Jiang2018field}%
\bibitem{Jiang2018doping}%
  \BibitemOpen
  \bibfield{author}{%
  \bibinfo {author} {\bibfnamefont{S.}~\bibnamefont{Jiang}}, \bibinfo {author}
  {\bibfnamefont{L.}~\bibnamefont{Li}}, \bibinfo {author}
  {\bibfnamefont{Z.}~\bibnamefont{Wang}}, \bibinfo {author}
  {\bibfnamefont{K.~F.}\ \bibnamefont{Mak}},\ and\ \bibinfo {author}
  {\bibfnamefont{J.}~\bibnamefont{Shan}},\ }%
  \Doi{10.1038/s41565-018-0135-x}{\emph{\bibinfo {title} {Controlling magnetism
  in {2D} {CrI$_3$} by electrostatic doping}}},\ \bibinfo {journal} {Nature
  Nanotechnology}\ \textbf{\bibinfo {volume} {13}},\ \bibinfo {pages} {549}
  (\bibinfo {year} {2018}).~%
  \bibAnnoteFile{Stop}{Jiang2018doping}%
\bibitem{Jiang2019}%
  \BibitemOpen
  \bibfield{author}{%
  \bibinfo {author} {\bibfnamefont{S.}~\bibnamefont{Jiang}}, \bibinfo {author}
  {\bibfnamefont{L.}~\bibnamefont{Li}}, \bibinfo {author}
  {\bibfnamefont{Z.}~\bibnamefont{Wang}}, \bibinfo {author}
  {\bibfnamefont{J.}~\bibnamefont{Shan}},\ and\ \bibinfo {author}
  {\bibfnamefont{K.~F.}\ \bibnamefont{Mak}},\ }%
  \Doi{10.1038/s41928-019-0232-3}{\emph{\bibinfo {title} {Spin tunnel
  field-effect transistors based on two-dimensional van der {Waals}
  heterostructures}}},\ \bibinfo {journal} {Nature Electronics}\
  \textbf{\bibinfo {volume} {2}},\ \bibinfo {pages} {159} (\bibinfo {year}
  {2019}).~%
  \bibAnnoteFile{Stop}{Jiang2019}%
\bibitem{Song2019}%
  \BibitemOpen
  \bibfield{author}{%
  \bibinfo {author} {\bibfnamefont{T.}~\bibnamefont{Song}}, \bibinfo {author}
  {\bibfnamefont{M.~W.-Y.}\ \bibnamefont{Tu}}, \bibinfo {author}
  {\bibfnamefont{C.}~\bibnamefont{Carnahan}}, \bibinfo {author}
  {\bibfnamefont{X.}~\bibnamefont{Cai}}, \bibinfo {author}
  {\bibfnamefont{T.}~\bibnamefont{Taniguchi}}, \bibinfo {author}
  {\bibfnamefont{K.}~\bibnamefont{Watanabe}}, \bibinfo {author}
  {\bibfnamefont{M.~A.}\ \bibnamefont{McGuire}}, \bibinfo {author}
  {\bibfnamefont{D.~H.}\ \bibnamefont{Cobden}}, \bibinfo {author}
  {\bibfnamefont{D.}~\bibnamefont{Xiao}}, \bibinfo {author}
  {\bibfnamefont{W.}~\bibnamefont{Yao}},\ and\ \bibinfo {author}
  {\bibfnamefont{X.}~\bibnamefont{Xu}},\ }%
  \Doi{10.1021/acs.nanolett.8b04160}{\emph{\bibinfo {title} {Voltage Control of
  a van der {Waals} Spin-Filter Magnetic Tunnel Junction}}},\ \bibinfo
  {journal} {Nano Letters}\ \textbf{\bibinfo {volume} {19}},\ \bibinfo {pages}
  {915} (\bibinfo {year} {2019}).~%
  \bibAnnoteFile{Stop}{Song2019}%
\bibitem{Djurdji2018}%
  \BibitemOpen
  \bibfield{author}{%
  \bibinfo {author} {\bibfnamefont{S.}~\bibnamefont{Djurdji\ifmmode
  \acute{c}\else~\'{c}\fi{} Mijin}}, \bibinfo {author}
  {\bibfnamefont{A.}~\bibnamefont{\ifmmode \check{S}\else
  \v{S}\fi{}olaji\ifmmode~\acute{c}\else \'{c}\fi{}}}, \bibinfo {author}
  {\bibfnamefont{J.}~\bibnamefont{Pe\ifmmode \check{s}\else
  \v{s}\fi{}i\ifmmode~\acute{c}\else \'{c}\fi{}}}, \bibinfo {author}
  {\bibfnamefont{M.}~\bibnamefont{\ifmmode \check{S}\else \v{S}\fi{}\ifmmode
  \acute{c}\else \'{c}\fi{}epanovi\ifmmode~\acute{c}\else \'{c}\fi{}}},
  \bibinfo {author} {\bibfnamefont{Y.}~\bibnamefont{Liu}}, \bibinfo {author}
  {\bibfnamefont{A.}~\bibnamefont{Baum}}, \bibinfo {author}
  {\bibfnamefont{C.}~\bibnamefont{Petrovic}}, \bibinfo {author}
  {\bibfnamefont{N.}~\bibnamefont{Lazarevi\ifmmode~\acute{c}\else
  \'{c}\fi{}}},\ and\ \bibinfo {author} {\bibfnamefont{Z.~V.}\
  \bibnamefont{Popovi\ifmmode~\acute{c}\else \'{c}\fi{}}},\ }%
  \Doi{10.1103/PhysRevB.98.104307}{\emph{\bibinfo {title} {Lattice dynamics and
  phase transition in ${\mathrm{CrI}}_{3}$ single crystals}}},\ \bibinfo
  {journal} {Phys. Rev. B}\ \textbf{\bibinfo {volume} {98}},\ \bibinfo {pages}
  {104307} (\bibinfo {year} {2018}).~%
  \bibAnnoteFile{Stop}{Djurdji2018}%
\bibitem{Sivadas2018}%
  \BibitemOpen
  \bibfield{author}{%
  \bibinfo {author} {\bibfnamefont{N.}~\bibnamefont{Sivadas}}, \bibinfo
  {author} {\bibfnamefont{S.}~\bibnamefont{Okamoto}}, \bibinfo {author}
  {\bibfnamefont{X.}~\bibnamefont{Xu}}, \bibinfo {author}
  {\bibfnamefont{C.~J.}\ \bibnamefont{Fennie}},\ and\ \bibinfo {author}
  {\bibfnamefont{D.}~\bibnamefont{Xiao}},\ }%
  \Doi{10.1021/acs.nanolett.8b03321}{\emph{\bibinfo {title} {Stacking-Dependent
  Magnetism in Bilayer CrI3}}},\ \bibinfo {journal} {Nano Letters}\
  \textbf{\bibinfo {volume} {18}},\ \bibinfo {pages} {7658} (\bibinfo {year}
  {2018}).~%
  \bibAnnoteFile{Stop}{Sivadas2018}%
\bibitem{Soriano2018}%
  \BibitemOpen
  \bibfield{author}{%
  \bibinfo {author} {\bibfnamefont{D.}~\bibnamefont{Soriano}}, \bibinfo
  {author} {\bibfnamefont{C.}~\bibnamefont{Cardoso}},\ and\ \bibinfo {author}
  {\bibfnamefont{J.}~\bibnamefont{Fern{\'a}ndez-Rossier}},\ }%
  \href{https://arxiv.org/abs/1807.00357}{\emph{\bibinfo {title} {Interplay
  between interlayer exchange and stacking in CrI$_3$}}},\ \bibinfo {journal}
  {arXiv:1807.00357}\  (\bibinfo {year} {2018}).~%
  \bibAnnoteFile{Stop}{Soriano2018}%
\bibitem{Jang2019}%
  \BibitemOpen
  \bibfield{author}{%
  \bibinfo {author} {\bibfnamefont{S.~W.}\ \bibnamefont{Jang}}, \bibinfo
  {author} {\bibfnamefont{M.~Y.}\ \bibnamefont{Jeong}}, \bibinfo {author}
  {\bibfnamefont{H.}~\bibnamefont{Yoon}}, \bibinfo {author}
  {\bibfnamefont{S.}~\bibnamefont{Ryee}},\ and\ \bibinfo {author}
  {\bibfnamefont{M.~J.}\ \bibnamefont{Han}},\ }%
  \Doi{10.1103/PhysRevMaterials.3.031001}{\emph{\bibinfo {title} {Microscopic
  understanding of magnetic interactions in bilayer ${\mathrm{CrI}}_{3}$}}},\
  \bibinfo {journal} {Phys. Rev. Materials}\ \textbf{\bibinfo {volume} {3}},\
  \bibinfo {pages} {031001} (\bibinfo {year} {2019}).~%
  \bibAnnoteFile{Stop}{Jang2019}%
\bibitem{Jiang2019theory}%
  \BibitemOpen
  \bibfield{author}{%
  \bibinfo {author} {\bibfnamefont{P.}~\bibnamefont{Jiang}}, \bibinfo {author}
  {\bibfnamefont{C.}~\bibnamefont{Wang}}, \bibinfo {author}
  {\bibfnamefont{D.}~\bibnamefont{Chen}}, \bibinfo {author}
  {\bibfnamefont{Z.}~\bibnamefont{Zhong}}, \bibinfo {author}
  {\bibfnamefont{Z.}~\bibnamefont{Yuan}}, \bibinfo {author}
  {\bibfnamefont{Z.-Y.}\ \bibnamefont{Lu}},\ and\ \bibinfo {author}
  {\bibfnamefont{W.}~\bibnamefont{Ji}},\ }%
  \Doi{10.1103/PhysRevB.99.144401}{\emph{\bibinfo {title} {Stacking tunable
  interlayer magnetism in bilayer ${\mathrm{CrI}}_{3}$}}},\ \bibinfo {journal}
  {Phys. Rev. B}\ \textbf{\bibinfo {volume} {99}},\ \bibinfo {pages} {144401}
  (\bibinfo {year} {2019}).~%
  \bibAnnoteFile{Stop}{Jiang2019theory}%
\bibitem{Lei2019}%
  \BibitemOpen
  \bibfield{author}{%
  \bibinfo {author} {\bibfnamefont{C.}~\bibnamefont{Lei}}, \bibinfo {author}
  {\bibfnamefont{B.~L.}\ \bibnamefont{Chittari}}, \bibinfo {author}
  {\bibfnamefont{K.}~\bibnamefont{Nomura}}, \bibinfo {author}
  {\bibfnamefont{N.}~\bibnamefont{Banerjee}}, \bibinfo {author}
  {\bibfnamefont{J.}~\bibnamefont{Jung}},\ and\ \bibinfo {author}
  {\bibfnamefont{A.~H.}\ \bibnamefont{MacDonald}},\ }%
  \href{https://arxiv.org/abs/1902.06418}{\emph{\bibinfo {title}
  {Magnetoelectric Response of Antiferromagnetic Van der Waals Bilayers}}},\
  \bibinfo {journal} {arXiv:1902.06418}\  (\bibinfo {year} {2019}).~%
  \bibAnnoteFile{Stop}{Lei2019}%
\bibitem{Song2019pressure}%
  \BibitemOpen
  \bibfield{author}{%
  \bibinfo {author} {\bibfnamefont{T.}~\bibnamefont{Song}}, \bibinfo {author}
  {\bibfnamefont{Z.}~\bibnamefont{Fei}}, \bibinfo {author}
  {\bibfnamefont{M.}~\bibnamefont{Yankowitz}}, \bibinfo {author}
  {\bibfnamefont{Z.}~\bibnamefont{Lin}}, \bibinfo {author}
  {\bibfnamefont{Q.}~\bibnamefont{Jiang}}, \bibinfo {author}
  {\bibfnamefont{K.}~\bibnamefont{Hwangbo}}, \bibinfo {author}
  {\bibfnamefont{Q.}~\bibnamefont{Zhang}}, \bibinfo {author}
  {\bibfnamefont{B.}~\bibnamefont{Sun}}, \bibinfo {author}
  {\bibfnamefont{T.}~\bibnamefont{Taniguchi}}, \bibinfo {author}
  {\bibfnamefont{K.}~\bibnamefont{Watanabe}}, \bibinfo {author}
  {\bibfnamefont{M.~A.}\ \bibnamefont{McGuire}}, \bibinfo {author}
  {\bibfnamefont{D.}~\bibnamefont{Graf}}, \bibinfo {author}
  {\bibfnamefont{T.}~\bibnamefont{Cao}}, \bibinfo {author}
  {\bibfnamefont{J.-H.}\ \bibnamefont{Chu}}, \bibinfo {author}
  {\bibfnamefont{D.~H.}\ \bibnamefont{Cobden}}, \bibinfo {author}
  {\bibfnamefont{C.~R.}\ \bibnamefont{Dean}}, \bibinfo {author}
  {\bibfnamefont{D.}~\bibnamefont{Xiao}},\ and\ \bibinfo {author}
  {\bibfnamefont{X.}~\bibnamefont{Xu}},\ }%
  \href{https://arxiv.org/abs/1905.10860}{\emph{\bibinfo {title} {Switching 2D
  Magnetic States via Pressure Tuning of Layer Stacking}}},\ \bibinfo {journal}
  {arXiv:1905.10860}\  (\bibinfo {year} {2019}).~%
  \bibAnnoteFile{Stop}{Song2019pressure}%
\bibitem{Li2019}%
  \BibitemOpen
  \bibfield{author}{%
  \bibinfo {author} {\bibfnamefont{T.}~\bibnamefont{Li}}, \bibinfo {author}
  {\bibfnamefont{S.}~\bibnamefont{Jiang}}, \bibinfo {author}
  {\bibfnamefont{N.}~\bibnamefont{Sivadas}}, \bibinfo {author}
  {\bibfnamefont{Z.}~\bibnamefont{Wang}}, \bibinfo {author}
  {\bibfnamefont{Y.}~\bibnamefont{Xu}}, \bibinfo {author}
  {\bibfnamefont{D.}~\bibnamefont{Weber}}, \bibinfo {author}
  {\bibfnamefont{J.~E.}\ \bibnamefont{Goldberger}}, \bibinfo {author}
  {\bibfnamefont{K.}~\bibnamefont{Watanabe}}, \bibinfo {author}
  {\bibfnamefont{T.}~\bibnamefont{Taniguchi}}, \bibinfo {author}
  {\bibfnamefont{C.~J.}\ \bibnamefont{Fennie}}, \bibinfo {author}
  {\bibfnamefont{K.~F.}\ \bibnamefont{Mak}},\ and\ \bibinfo {author}
  {\bibfnamefont{J.}~\bibnamefont{Shan}},\ }%
  \href{https://arxiv.org/abs/1905.10905}{\emph{\bibinfo {title}
  {Pressure-controlled interlayer magnetism in atomically thin {CrI}$_3$}}},\
  \bibinfo {journal} {arXiv:1905.10905}\  (\bibinfo {year} {2019}).~%
  \bibAnnoteFile{Stop}{Li2019}%
\bibitem{Chen2019}%
  \BibitemOpen
  \bibfield{author}{%
  \bibinfo {author} {\bibfnamefont{W.}~\bibnamefont{Chen}}, \bibinfo {author}
  {\bibfnamefont{Z.}~\bibnamefont{Sun}}, \bibinfo {author}
  {\bibfnamefont{L.}~\bibnamefont{Gu}}, \bibinfo {author}
  {\bibfnamefont{X.}~\bibnamefont{Xu}}, \bibinfo {author}
  {\bibfnamefont{S.}~\bibnamefont{Wu}},\ and\ \bibinfo {author}
  {\bibfnamefont{C.}~\bibnamefont{Gao}},\ }%
  \href{https://arxiv.org/abs/1906.03383}{\emph{\bibinfo {title} {Direct
  observation of van der Waals stacking dependent interlayer magnetism}}},\
  \bibinfo {journal} {arXiv:1906.03383}\  (\bibinfo {year} {2019}).~%
  \bibAnnoteFile{Stop}{Chen2019}%
\bibitem{Sun2019}%
  \BibitemOpen
  \bibfield{author}{%
  \bibinfo {author} {\bibfnamefont{Z.}~\bibnamefont{Sun}}, \bibinfo {author}
  {\bibfnamefont{Y.}~\bibnamefont{Yi}}, \bibinfo {author}
  {\bibfnamefont{T.}~\bibnamefont{Song}}, \bibinfo {author}
  {\bibfnamefont{G.}~\bibnamefont{Clark}}, \bibinfo {author}
  {\bibfnamefont{B.}~\bibnamefont{Huang}}, \bibinfo {author}
  {\bibfnamefont{Y.}~\bibnamefont{Shan}}, \bibinfo {author}
  {\bibfnamefont{S.}~\bibnamefont{Wu}}, \bibinfo {author}
  {\bibfnamefont{D.}~\bibnamefont{Huang}}, \bibinfo {author}
  {\bibfnamefont{C.}~\bibnamefont{Gao}}, \bibinfo {author}
  {\bibfnamefont{Z.}~\bibnamefont{Chen}}, \bibinfo {author}
  {\bibfnamefont{M.}~\bibnamefont{McGuire}}, \bibinfo {author}
  {\bibfnamefont{T.}~\bibnamefont{Cao}}, \bibinfo {author}
  {\bibfnamefont{D.}~\bibnamefont{Xiao}}, \bibinfo {author}
  {\bibfnamefont{W.-T.}\ \bibnamefont{Liu}}, \bibinfo {author}
  {\bibfnamefont{W.}~\bibnamefont{Yao}}, \bibinfo {author}
  {\bibfnamefont{X.}~\bibnamefont{Xu}},\ and\ \bibinfo {author}
  {\bibfnamefont{S.}~\bibnamefont{Wu}},\ }%
  \href{https://arxiv.org/abs/1904.03577}{\emph{\bibinfo {title} {Giant and
  nonreciprocal second harmonic generation from layered antiferromagnetism in
  bilayer CrI3}}},\ \bibinfo {journal} {arXiv:1904.03577}\  (\bibinfo {year}
  {2019}).~%
  \bibAnnoteFile{Stop}{Sun2019}%
\bibitem{Klein2019}%
  \BibitemOpen
  \bibfield{author}{%
  \bibinfo {author} {\bibfnamefont{D.~R.}\ \bibnamefont{Klein}}, \bibinfo
  {author} {\bibfnamefont{D.}~\bibnamefont{MacNeill}}, \bibinfo {author}
  {\bibfnamefont{Q.}~\bibnamefont{Song}}, \bibinfo {author}
  {\bibfnamefont{D.~T.}\ \bibnamefont{Larson}}, \bibinfo {author}
  {\bibfnamefont{S.}~\bibnamefont{Fang}}, \bibinfo {author}
  {\bibfnamefont{M.}~\bibnamefont{Xu}}, \bibinfo {author}
  {\bibfnamefont{R.~A.}\ \bibnamefont{Ribeiro}}, \bibinfo {author}
  {\bibfnamefont{P.~C.}\ \bibnamefont{Canfield}}, \bibinfo {author}
  {\bibfnamefont{E.}~\bibnamefont{Kaxiras}}, \bibinfo {author}
  {\bibfnamefont{R.}~\bibnamefont{Comin}},\ and\ \bibinfo {author}
  {\bibfnamefont{P.}~\bibnamefont{Jarillo-Herrero}},\ }%
  \href{https://arxiv.org/abs/1903.00002}{\emph{\bibinfo {title} {Giant
  enhancement of interlayer exchange in an ultrathin {2D} magnet}}},\ \bibinfo
  {journal} {arXiv:1903.00002}\  (\bibinfo {year} {2019}).~%
  \bibAnnoteFile{Stop}{Klein2019}%
\bibitem{Giannozzi2009}%
  \BibitemOpen
  \bibfield{author}{%
  \bibinfo {author} {\bibfnamefont{P.}~\bibnamefont{Giannozzi}}, \bibinfo
  {author} {\bibfnamefont{S.}~\bibnamefont{Baroni}}, \bibinfo {author}
  {\bibfnamefont{N.}~\bibnamefont{Bonini}}, \bibinfo {author}
  {\bibfnamefont{M.}~\bibnamefont{Calandra}}, \bibinfo {author}
  {\bibfnamefont{R.}~\bibnamefont{Car}}, \bibinfo {author}
  {\bibfnamefont{C.}~\bibnamefont{Cavazzoni}}, \bibinfo {author}
  {\bibfnamefont{D.}~\bibnamefont{Ceresoli}}, \bibinfo {author}
  {\bibfnamefont{G.~L.}\ \bibnamefont{Chiarotti}}, \bibinfo {author}
  {\bibfnamefont{M.}~\bibnamefont{Cococcioni}}, \bibinfo {author}
  {\bibfnamefont{I.}~\bibnamefont{Dabo}}, \bibinfo {author}
  {\bibfnamefont{A.}~\bibnamefont{{Dal Corso}}}, \bibinfo {author}
  {\bibfnamefont{S.}~\bibnamefont{de~Gironcoli}}, \bibinfo {author}
  {\bibfnamefont{S.}~\bibnamefont{Fabris}}, \bibinfo {author}
  {\bibfnamefont{G.}~\bibnamefont{Fratesi}}, \bibinfo {author}
  {\bibfnamefont{R.}~\bibnamefont{Gebauer}}, \bibinfo {author}
  {\bibfnamefont{U.}~\bibnamefont{Gerstmann}}, \bibinfo {author}
  {\bibfnamefont{C.}~\bibnamefont{Gougoussis}}, \bibinfo {author}
  {\bibfnamefont{A.}~\bibnamefont{Kokalj}}, \bibinfo {author}
  {\bibfnamefont{M.}~\bibnamefont{Lazzeri}}, \bibinfo {author}
  {\bibfnamefont{L.}~\bibnamefont{Martin-Samos}}, \bibinfo {author}
  {\bibfnamefont{N.}~\bibnamefont{Marzari}}, \bibinfo {author}
  {\bibfnamefont{F.}~\bibnamefont{Mauri}}, \bibinfo {author}
  {\bibfnamefont{R.}~\bibnamefont{Mazzarello}}, \bibinfo {author}
  {\bibfnamefont{S.}~\bibnamefont{Paolini}}, \bibinfo {author}
  {\bibfnamefont{A.}~\bibnamefont{Pasquarello}}, \bibinfo {author}
  {\bibfnamefont{L.}~\bibnamefont{Paulatto}}, \bibinfo {author}
  {\bibfnamefont{C.}~\bibnamefont{Sbraccia}}, \bibinfo {author}
  {\bibfnamefont{S.}~\bibnamefont{Scandolo}}, \bibinfo {author}
  {\bibfnamefont{G.}~\bibnamefont{Sclauzero}}, \bibinfo {author}
  {\bibfnamefont{A.~P.}\ \bibnamefont{Seitsonen}}, \bibinfo {author}
  {\bibfnamefont{A.}~\bibnamefont{Smogunov}}, \bibinfo {author}
  {\bibfnamefont{P.}~\bibnamefont{Umari}},\ and\ \bibinfo {author}
  {\bibfnamefont{R.~M.}\ \bibnamefont{Wentzcovitch}},\ }%
  \Doi{10.1088/0953-8984/21/39/395502}{\emph{\bibinfo {title} {{QUANTUM
  ESPRESSO: a modular and open-source software project for quantum simulations
  of materials.}}}},\ \bibinfo {journal} {Journal of Physics: Condensed
  Matter}\ \textbf{\bibinfo {volume} {21}},\ \bibinfo {pages} {395502}
  (\bibinfo {year} {2009}).~%
  \bibAnnoteFile{Stop}{Giannozzi2009}%
\bibitem{Giannozzi2017}%
  \BibitemOpen
  \bibfield{author}{%
  \bibinfo {author} {\bibfnamefont{P.}~\bibnamefont{Giannozzi}}, \bibinfo
  {author} {\bibfnamefont{O.}~\bibnamefont{Andreussi}}, \bibinfo {author}
  {\bibfnamefont{T.}~\bibnamefont{Brumme}}, \bibinfo {author}
  {\bibfnamefont{O.}~\bibnamefont{Bunau}}, \bibinfo {author}
  {\bibfnamefont{M.}~\bibnamefont{{Buongiorno Nardelli}}}, \bibinfo {author}
  {\bibfnamefont{M.}~\bibnamefont{Calandra}}, \bibinfo {author}
  {\bibfnamefont{R.}~\bibnamefont{Car}}, \bibinfo {author}
  {\bibfnamefont{C.}~\bibnamefont{Cavazzoni}}, \bibinfo {author}
  {\bibfnamefont{D.}~\bibnamefont{Ceresoli}}, \bibinfo {author}
  {\bibfnamefont{M.}~\bibnamefont{Cococcioni}}, \bibinfo {author}
  {\bibfnamefont{N.}~\bibnamefont{Colonna}}, \bibinfo {author}
  {\bibfnamefont{I.}~\bibnamefont{Carnimeo}}, \bibinfo {author}
  {\bibfnamefont{A.}~\bibnamefont{{Dal Corso}}}, \bibinfo {author}
  {\bibfnamefont{S.}~\bibnamefont{de~Gironcoli}}, \bibinfo {author}
  {\bibfnamefont{P.}~\bibnamefont{Delugas}}, \bibinfo {author}
  {\bibfnamefont{R.~A.}\ \bibnamefont{DiStasio}}, \bibinfo {author}
  {\bibfnamefont{A.}~\bibnamefont{Ferretti}}, \bibinfo {author}
  {\bibfnamefont{A.}~\bibnamefont{Floris}}, \bibinfo {author}
  {\bibfnamefont{G.}~\bibnamefont{Fratesi}}, \bibinfo {author}
  {\bibfnamefont{G.}~\bibnamefont{Fugallo}}, \bibinfo {author}
  {\bibfnamefont{R.}~\bibnamefont{Gebauer}}, \bibinfo {author}
  {\bibfnamefont{U.}~\bibnamefont{Gerstmann}}, \bibinfo {author}
  {\bibfnamefont{F.}~\bibnamefont{Giustino}}, \bibinfo {author}
  {\bibfnamefont{T.}~\bibnamefont{Gorni}}, \bibinfo {author}
  {\bibfnamefont{J.}~\bibnamefont{Jia}}, \bibinfo {author}
  {\bibfnamefont{M.}~\bibnamefont{Kawamura}}, \bibinfo {author}
  {\bibfnamefont{H.-Y.}\ \bibnamefont{Ko}}, \bibinfo {author}
  {\bibfnamefont{A.}~\bibnamefont{Kokalj}}, \bibinfo {author}
  {\bibfnamefont{E.}~\bibnamefont{K{\"{u}}{\c{c}}{\"{u}}kbenli}}, \bibinfo
  {author} {\bibfnamefont{M.}~\bibnamefont{Lazzeri}}, \bibinfo {author}
  {\bibfnamefont{M.}~\bibnamefont{Marsili}}, \bibinfo {author}
  {\bibfnamefont{N.}~\bibnamefont{Marzari}}, \bibinfo {author}
  {\bibfnamefont{F.}~\bibnamefont{Mauri}}, \bibinfo {author}
  {\bibfnamefont{N.~L.}\ \bibnamefont{Nguyen}}, \bibinfo {author}
  {\bibfnamefont{H.-V.}\ \bibnamefont{Nguyen}}, \bibinfo {author}
  {\bibfnamefont{A.}~\bibnamefont{Otero-de-la Roza}}, \bibinfo {author}
  {\bibfnamefont{L.}~\bibnamefont{Paulatto}}, \bibinfo {author}
  {\bibfnamefont{S.}~\bibnamefont{Ponc{\'{e}}}}, \bibinfo {author}
  {\bibfnamefont{D.}~\bibnamefont{Rocca}}, \bibinfo {author}
  {\bibfnamefont{R.}~\bibnamefont{Sabatini}}, \bibinfo {author}
  {\bibfnamefont{B.}~\bibnamefont{Santra}}, \bibinfo {author}
  {\bibfnamefont{M.}~\bibnamefont{Schlipf}}, \bibinfo {author}
  {\bibfnamefont{A.~P.}\ \bibnamefont{Seitsonen}}, \bibinfo {author}
  {\bibfnamefont{A.}~\bibnamefont{Smogunov}}, \bibinfo {author}
  {\bibfnamefont{I.}~\bibnamefont{Timrov}}, \bibinfo {author}
  {\bibfnamefont{T.}~\bibnamefont{Thonhauser}}, \bibinfo {author}
  {\bibfnamefont{P.}~\bibnamefont{Umari}}, \bibinfo {author}
  {\bibfnamefont{N.}~\bibnamefont{Vast}}, \bibinfo {author}
  {\bibfnamefont{X.}~\bibnamefont{Wu}},\ and\ \bibinfo {author}
  {\bibfnamefont{S.}~\bibnamefont{Baroni}},\ }%
  \Doi{10.1088/1361-648X/aa8f79}{\emph{\bibinfo {title} {{Advanced capabilities
  for materials modelling with Quantum ESPRESSO}}}},\ \bibinfo {journal}
  {Journal of Physics: Condensed Matter}\ \textbf{\bibinfo {volume} {29}},\
  \bibinfo {pages} {465901} (\bibinfo {year} {2017}).~%
  \bibAnnoteFile{Stop}{Giannozzi2017}%
\bibitem{Thonhauser2015}%
  \BibitemOpen
  \bibfield{author}{%
  \bibinfo {author} {\bibfnamefont{T.}~\bibnamefont{Thonhauser}}, \bibinfo
  {author} {\bibfnamefont{S.}~\bibnamefont{Zuluaga}}, \bibinfo {author}
  {\bibfnamefont{C.~A.}\ \bibnamefont{Arter}}, \bibinfo {author}
  {\bibfnamefont{K.}~\bibnamefont{Berland}}, \bibinfo {author}
  {\bibfnamefont{E.}~\bibnamefont{Schr\"oder}},\ and\ \bibinfo {author}
  {\bibfnamefont{P.}~\bibnamefont{Hyldgaard}},\ }%
  \Doi{10.1103/PhysRevLett.115.136402}{\emph{\bibinfo {title} {Spin Signature
  of Nonlocal Correlation Binding in Metal-Organic Frameworks}}},\ \bibinfo
  {journal} {Phys. Rev. Lett.}\ \textbf{\bibinfo {volume} {115}},\ \bibinfo
  {pages} {136402} (\bibinfo {year} {2015}).~%
  \bibAnnoteFile{Stop}{Thonhauser2015}%
\bibitem{Dion2004}%
  \BibitemOpen
  \bibfield{author}{%
  \bibinfo {author} {\bibfnamefont{M.}~\bibnamefont{Dion}}, \bibinfo {author}
  {\bibfnamefont{H.}~\bibnamefont{Rydberg}}, \bibinfo {author}
  {\bibfnamefont{E.}~\bibnamefont{Schr\"oder}}, \bibinfo {author}
  {\bibfnamefont{D.~C.}\ \bibnamefont{Langreth}},\ and\ \bibinfo {author}
  {\bibfnamefont{B.~I.}\ \bibnamefont{Lundqvist}},\ }%
  \Doi{10.1103/PhysRevLett.92.246401}{\emph{\bibinfo {title} {Van der Waals
  Density Functional for General Geometries}}},\ \bibinfo {journal} {Phys. Rev.
  Lett.}\ \textbf{\bibinfo {volume} {92}},\ \bibinfo {pages} {246401} (\bibinfo
  {year} {2004}).~%
  \bibAnnoteFile{Stop}{Dion2004}%
\bibitem{Lee2010}%
  \BibitemOpen
  \bibfield{author}{%
  \bibinfo {author} {\bibfnamefont{K.}~\bibnamefont{Lee}}, \bibinfo {author}
  {\bibfnamefont{E.~D.}\ \bibnamefont{Murray}}, \bibinfo {author}
  {\bibfnamefont{L.}~\bibnamefont{Kong}}, \bibinfo {author}
  {\bibfnamefont{B.~I.}\ \bibnamefont{Lundqvist}},\ and\ \bibinfo {author}
  {\bibfnamefont{D.~C.}\ \bibnamefont{Langreth}},\ }%
  \Doi{10.1103/PhysRevB.82.081101}{\emph{\bibinfo {title} {Higher-accuracy van
  der Waals density functional}}},\ \bibinfo {journal} {Phys. Rev. B}\
  \textbf{\bibinfo {volume} {82}},\ \bibinfo {pages} {081101} (\bibinfo {year}
  {2010}).~%
  \bibAnnoteFile{Stop}{Lee2010}%
\bibitem{Togo2015}%
  \BibitemOpen
  \bibfield{author}{%
  \bibinfo {author} {\bibfnamefont{A.}~\bibnamefont{Togo}}\ and\ \bibinfo
  {author} {\bibfnamefont{I.}~\bibnamefont{Tanaka}},\ }%
  \Doi{10.1016/j.scriptamat.2015.07.021}{\emph{\bibinfo {title} {First
  principles phonon calculations in materials science}}},\ \bibinfo {journal}
  {Scr. Mater.}\ \textbf{\bibinfo {volume} {108}},\ \bibinfo {pages} {1}
  (\bibinfo {year} {2015}).~%
  \bibAnnoteFile{Stop}{Togo2015}%
\bibitem{Prandini2018}%
  \BibitemOpen
  \bibfield{author}{%
  \bibinfo {author} {\bibfnamefont{G.}~\bibnamefont{Prandini}}, \bibinfo
  {author} {\bibfnamefont{A.}~\bibnamefont{Marrazzo}}, \bibinfo {author}
  {\bibfnamefont{I.~E.}\ \bibnamefont{Castelli}}, \bibinfo {author}
  {\bibfnamefont{N.}~\bibnamefont{Mounet}},\ and\ \bibinfo {author}
  {\bibfnamefont{N.}~\bibnamefont{Marzari}},\ }%
  \Doi{10.1038/s41524-018-0127-2}{\emph{\bibinfo {title} {Precision and
  efficiency in solid-state pseudopotential calculations}}},\ \bibinfo
  {journal} {npj Computational Materials}\ \textbf{\bibinfo {volume} {4}},\
  \bibinfo {pages} {72} (\bibinfo {year} {2018}).~%
  \bibAnnoteFile{Stop}{Prandini2018}%
\bibitem{mcgrane_quantitative_2014}%
  \BibitemOpen
  \bibfield{author}{%
  \bibinfo {author} {\bibfnamefont{S.~D.}\ \bibnamefont{McGrane}}, \bibinfo
  {author} {\bibfnamefont{D.~S.}\ \bibnamefont{Moore}}, \bibinfo {author}
  {\bibfnamefont{P.~M.}\ \bibnamefont{Goodwin}},\ and\ \bibinfo {author}
  {\bibfnamefont{D.~M.}\ \bibnamefont{Dattelbaum}},\ }%
  \Doi{10.1366/14-07503}{\emph{\bibinfo {title} {Quantitative {Tradeoffs}
  between {Spatial}, {Temporal}, and {Thermometric} {Resolution} of
  {Nonresonant} {Raman} {Thermometry} for {Dynamic} {Experiments}}}},\ \bibinfo
  {journal} {Applied Spectroscopy}\ \textbf{\bibinfo {volume} {68}},\ \bibinfo
  {pages} {1279} (\bibinfo {year} {2014}).~%
  \bibAnnoteFile{Stop}{mcgrane_quantitative_2014}%
\end{thebibliography}
\end{document}